\documentclass[letter]{aa}  

\usepackage{graphicx}
\usepackage{txfonts}
\usepackage{float}

\usepackage{orcidlink}
\newcommand{\kms}{$\mathrm{km~s^{-1}}$}
\newcommand{\kmskpc}{$\mathrm{km~s^{-1}\,kpc^{-1}}$}
\newcommand{\omp}{$\Omega_p$}

\newcommand{\gaia}{\textit{Gaia}}
\defcitealias{2023jimenez}{JA23}
\defcitealias{2024ajimenez}{JA24}
\defcitealias{2021luri}{GCL21}

\begin{document} 
   \title{Bar pattern speed modulation across LMC stellar populations}
   \author{V. Araya\inst{1}\,\orcidlink{0009-0008-6029-7201}
          \and
          L. Chemin\inst{2}\,\orcidlink{0000-0002-3834-7937}
          \and
          \'O. Jim\'enez-Arranz\inst{3}\,\orcidlink{0000-0001-7434-5165}
          \and 
          M. Romero-G\'omez \inst{4,5,6}\,\orcidlink{0000-0003-3936-1025}
          }
   \institute{Instituto de Astrof\'isica, Departamento de F\'isica y Astronomía,
Facultad de Ciencias Exactas, Universidad Andr\'es Bello, Fernandez Concha 700, Las Condes, Santiago RM, Chile,  \email{v.arayacampos@uandresbello.edu}
         \and
             Université de Strasbourg, CNRS, Observatoire Astronomique de Strasbourg, UMR 7550, 67000 Strasbourg, France, \email{laurent.chemin@unistra.fr}
        \and
            Lund Observatory, Division of Astrophysics, Lund University, Box 43, 221 00 Lund, Sweden
        \and
        Departament de Física Quàntica i Astrofísica (FQA), Universitat de Barcelona (UB), C Martí i Franquès, 1, 08028 Barcelona, Spain
        \and
        Institut de Ciències del Cosmos (ICCUB), Universitat de Barcelona, Martí i Franquès 1, 08028 Barcelona, Spain
        \and
        Institut d’Estudis Espacials de Catalunya (IEEC), c. Esteve Terradas 1, 08860 Castelldefels (Barcelona), Spain 
        }
   \date{Received month dd, yyyy; accepted month dd, yyyy}

  \abstract
   {The bar pattern speed of the LMC has been measured using \gaia\ data, suggesting the presence of a slow pattern, perhaps not rotating at all. Numerical simulations of interacting LMC-SMC systems were able to reproduce a bar stoppage.
   Here, we report on the first measurement of the bar pattern speed of the LMC as a function of the evolutionary phase of its stellar populations. 
   Astrometric and photometric data of $\sim$11 million LMC stars from  \gaia\ DR3 were used to build five evolutionary phases, from less to more evolved stars. The Dehnen method, a new procedure to derive bar pattern speeds from kinematics of particles in N-body simulations, is applied to the LMC stellar populations. 
   We observe a modulation of the bar pattern speed with the evolutionary phase,  meaning that different LMC stellar populations exhibit  different pattern speeds, ranging  from $-0.9$ to 6.6 \kmskpc. Moreover, less evolved stars have a lower pattern speed while the bar of more evolved phases tends to rotate faster.  The LMC bar is thus extremely slow, ruling out the presence of bar corotation within the disc, in agreement with a previous claim, but this time observed with various stellar populations. 
   It is the first time that a pattern speed is measured separately for different stellar populations in any galaxy. The LMC pattern speed cannot be simply resumed to a singular value, but instead is an overlay of different patterns depending on the evolutionary phase of the stars. Future \gaia\ releases will be crucial to investigate more deeply the relations of the pattern speed with important astrophysical parameters of stars, like their age and metallicity, which will be helpful to constrain the chemo-dynamical evolution of the LMC bar.}

\titlerunning{Bar pattern speed modulation across LMC stellar populations}

\maketitle

\section{Introduction}
\label{sec:intro}

Bars in galactic discs are long-lived features which have a crucial role in the secular evolution of the structure and dynamics of stars and gas  \citep{2025fragkoudi}. A fundamental property of a bar is its pattern speed \omp, which is a key parameter from which stellar orbits and their relations with the  bar-disc resonances  can be studied. Only a couple of hundreds of bar pattern speeds of galaxies have been measured so far \citep[e.g.][]{2020cuomo,2021williams,2023geron}, almost exclusively by means of the Tremaine-Weinberg method \citep[][hereafter TW]{1984tw} applied to stellar line-of-sight velocity fields. In the TW method, \omp\ is deduced from a linear relation between  integrals of velocities and positions measured along directions parallel to the line-of-nodes.

The pioneer work to use both astrometric and spectroscopic data of stars to infer a bar \omp\  for a galaxy other than the Milky Way is that of \citet[][hereafter JA24]{2024ajimenez}. These authors measured \omp\ of the stellar bar of the Large Magellanic Cloud (LMC) using  the Third Data Release of the \gaia\ mission \citep{2016prusti,2021abrown,2021bbrown,2023vallenari}. Their analysis is very valuable in many aspects. 
First, since \gaia\ proper motions allow one to get LMC disc velocity fields in a Cartesian frame with about 11 million stars, it was possible to apply the TW method in its original form, called ``in-plane TW method'' (or IPTW, as opposed to the traditional line-of-sight velocity prescription, or LTW). The advantage of applying the TW method in the Cartesian frame is that we can place the $x-y$ plane at any arbitrary position around the $z-$axis perpendicular to the disc before performing the integrals of positions and velocities. Therefore, by rotating the $x-y$ plane around the $z-$axis, there is no limit to the number of values of \omp\ that can be measured, making it possible to study its variation as a function of the frame orientation, and likewise, of the bar orientation with respect to the main axes of the disc. \citetalias{2024ajimenez} showed that the IPTW method yields a wide range of LMC bar pattern speeds ($0-60$ \kmskpc), owing to the strong variation of the integrals with the orientation of the frame. Variations occur at any azimuthal angle, not only near the critical minor and major axes of the stellar bar.  In addition, 
they applied the LTW method to $\sim 30\,000$ line-of-sight velocities measured by the \gaia\ Radial Velocity Spectrometer 
\citep[RVS,][]{2018cropper,2023katz}, and found $\Omega_p \sim 30$ \kmskpc, which corresponds exactly to the IPTW value when the $x-$axis is aligned with the line-of-nodes. The agreement between the IPTW and LTW values at this exact orientation is evidence that the LTW value is only representative of a particular orientation of the bar, not of the real \omp. In other words,  it is highly probable that the LTW method would have yielded any value other than 30 \kmskpc, within $0-60$ \kmskpc, had the LMC bar been oriented differently with respect to the line-of-nodes. Note that LMC spiral arms could contribute to the issues faced by the TW method as well. 

Second, the failure of the TW method to yield a unique pattern speed prompted \citetalias{2024ajimenez} to employ other independent methods. One of them  searches for the corotation radius $R_c$ near the location where the phase angle of the second-order Fourier mode in the tangential velocity field varies by $\sim$90\degr\ (the ``BV model''). Indeed, inside $R_c$, the periodic stable orbits are aligned with the bar potential \citep{1980contopoulos,1983athanassoula,1989contopoulos} and the tangential velocity is minimum (maximum, respectively) along the bar major axis (perpendicularly),  while beyond $R_c$, it is the opposite; hence stellar orbits no longer construct an elongated pattern beyond corotation. Note that these theoretical and numerical considerations are additional evidence that \omp\ as fast as, e.g., 30 \kmskpc\ as measured with the LTW method must be ruled out because it would correspond to an unphysical ultra fast bar with a radius 2.3 times larger than the corresponding corotation radius ($R_c \sim 1$ kpc, see Fig. 11 of \citetalias{2024ajimenez}).
\citetalias{2024ajimenez} thus fitted the BV model to the LMC tangential velocity field at low radius, and found $R_c = 4.2$ kpc,   corresponding to a slow bar with $\Omega_p = 18.5^{+1.2}_{-1.1}$ \kmskpc. Unlike the TW method, the BV model has no  mathematical basis. Also, though seemingly promising with simulations, its application to  data remains limited  -- only galaxies with resolved planar velocities can be studied that way, i.e. only the LMC and the Galaxy \citep{2023drimmel} -- and uncertain  because it assumes that the variation of the kinematic bisymmetry by $\sim$90\degr\ is due to the bar dynamics only. However, the LMC exhibits a spiral structure which develops along the leading side of the bar so that it is possible that $R_c=4.2$ kpc is a lower limit value of the bar corotation, and $\Omega_p = 18.5$ \kmskpc\ an upper limit, as spiral arms could begin to affect the searched variation of the phase angle at a too small radius. 

Third, even more compelling, \citetalias{2024ajimenez} applied  the new  mathematical model by \citet[][the ``Dehnen method'']{2023dehnen}, which was developed to infer   bar pattern speeds  from single snapshots of numerical simulations. Given that the method utilizes the density and planar kinematics of stellar-like particles, it should  be well-suited for studying the properties of the bars of the Milky Way \citep{2024zhang} and  the LMC   using \gaia\ astrometry of stars. \citetalias{2024ajimenez}  evidenced  an almost non-rotating  LMC bar, perhaps with a slight hint of counter-rotation ($\Omega_p = -1.0 \pm 0.5$ \kmskpc). This surprising result suggests the important role of the interaction with the Small Magellanic Cloud (SMC) on the dynamics of the LMC, even in the inner disc region. To assess this possibility, \citet{2025jimenez} then searched for non-rotating bars in KRATOS, a suite of numerical simulations of tidal LMC-SMC-like interacting systems \citep{2024bjimenez}, and found that bars can slow down until stopping their rotation. This process occurs rapidly after the second  passage   of the SMC-like companion at pericentre, followed by a temporary ($\sim 100$ Myr)  stage of extremely low rotation  until the \omp\ rises again. 

This stunning outcome in \citetalias{2024ajimenez} marks the first discovery of an almost stationary bar, and further analysis will undoubtedly be necessary to fully understand its implications for the evolution and dynamics of bars, as well as the recent evolution of the LMC. Within this framework, we propose   to extend the analysis of \citetalias{2024ajimenez}  to investigate for the first time the properties of the bar pattern speed as a function of the evolutionary phase of stellar populations in the LMC. How does \omp\ of less evolved stars in the bar compare with that of evolved stars? Is there a singular pattern speed for the LMC  bar? To explore these  questions, we leverage the selection of various stellar evolutionary phases in a  color-magnitude diagram (CMD) of LMC stars from \citet[][hereafter GCL21]{2021luri}, using the refined samples provided by \citet[][hereafter JA23]{2023jimenez} to measure their bar pattern speed by means of the Dehnen method.

\section{Methodology and Observational data}
\label{sec:methodobs}


We use the program \texttt{patternSpeed.py} (v0.2) of \citet{2023dehnen} to infer \omp\ and the phase angle  
$\phi_b$ of the bar. They are measured from the density and kinematics once the bar region, defined by its boundaries $[R_0, R_1]$, has been identified from a Fourier analysis of the stellar density. 
The bar angle is the phase angle  of the second-order Fourier coefficient of the density, and \omp\ is its time derivative 
$\mathrm{d}\phi_b/\mathrm{d}t$. More details are summarized in App.~\ref{sec:app1}.
The Dehnen method performs extremely well with single snapshots of numerical simulations of isolated or interacting barred spirals when compared to the direct measurement of the pattern speeds from differentials of $\phi_b$ along time consecutive snapshots  \citep[][]{2023dehnen,2024ajimenez,2024bjimenez,2024semczuk,2024zhang,2025jimenez}.  It remains  stable and robust, regardless of the dynamical equilibrium of the outer disc. Comparisons with the TW method applied to mock data of  discs simulated in a cosmological context for the case where the bar phase makes and angle of $45\degr$ with respect to the disc major axis is also performed in \citet{2025roshan}.
With the LMC data, the likes of particles are the stars. 

The data used in this study are from the \gaia\ mission \citep{2016prusti}, as provided by the third version of its catalogue \citep{2021abrown,2021bbrown,2023vallenari}. Our samples of stars at various evolutionary stages  are constructed from the sample of LMC stars defined in \citetalias{2023jimenez}, to which we applied the evolutionary phases described by \citetalias{2021luri}. 
\citetalias{2023jimenez} applied neural network (NN) classifications which provided different samples depending on the desired level of completeness/purity. They defined a sample which prioritizes completeness (hereafter NN Complete sample) made of $12\,116\, 762$ LMC stars, and which is our starting point. We build stellar sub-samples from it by applying the polygonial areas designed by \citetalias{2021luri} from a CMD of  LMC stars (see their Fig. 2), delineating eight evolutionary phases.    Main Sequence (MS) stars in \citetalias{2021luri} were  split in three sub-samples  spread over  stars as young as a few tens of Myr and as old as $\sim$2 Gyr.  They were referred to as ``young'' stellar populations, being less evolved and/or younger than Red Clump (RC), Red Giant Branch (RGB), Asymptotic Giant Branch (AGB) and RR Lyrae (RRL) stars. The RR Lyrae phase represents the oldest stars in their sample ($\gtrsim$10 Gyr). \citetalias{2021luri} also identified the Blue Loop (BL) sub-sample as  stars from the CMD mostly located closer to the evolved RGB, RC and AGB phases, and as the feature connecting to the young phase of the MS. The BL sub-sample is thus made of stars of various ages which depend on their initial masses. Only a few percent of them are classical Cepheids \citep{2022ripepi}, thus are as young as $50-350$ Myr (\citetalias{2021luri}) and as the younger stars of the MS though not being at the same stage of evolution, while other BL stars  can be as old as few Gyr (the least massive ones). 
For simplicity, we do not make any difference between the three evolutionary phases along the MS, and merge them into a  single phase representing a sub-sample of $4\,500\,453$ stars referred to as the ``Young'' sub-sample, by analogy to \citetalias{2021luri}. Moreover,  we merged the most evolved AGB and RRL stages of \citetalias{2021luri} into an ``AGB+RRL'' sub-sample with $278\, 629$ stars as they were individually sub-samples with smaller numbers of stars. Our RC, RGB and BL sub-samples are made of $3\,737\,496$, $2\,765\,725$ and $252\,958$ stars, respectively.  In total, our final sample is thus made of $11\,535\,261$ stars  spread over five evolutionary phases (Tab.~\ref{tab:res}). The difference of $581\,501$ stars with the NN Complete sample of \citetalias{2023jimenez} and \citetalias{2024ajimenez} is due to stars  located in a few areas of the CMD which could not be assigned to any particular phases (see more details in \citetalias{2021luri}). 
We assume hereafter the same LMC parameters as in \citetalias{2021luri} and \citetalias{2024ajimenez} (and reference therein):  distance of 49.59 kpc \citep{2019pietrzynski}, equatorial coordinates of the kinematic centre at $(81.28\degr,-69.78\degr)$,  disc inclination ($34\degr$) and position angle of the major axis ($310\degr$). These values were derived by fitting an infinitely thin plane of constant inclination and position angle, assuming axisymmetric rotational motions (\citetalias{2021luri}). We thus implicitely assume that all stars lie in the plane of the LMC. It is nonetheless worth notifying that  the inner and outer parts of the LMC disc may not be fully coplanar owing to an asymmetric disc warping \citep{2025bjimenez}. Additionally,  given that distances to some LMC stars can be constrained with methods such as the Tip of the RGB, RC stars \citep{2022saroon}, or variable stars like Cepheids \citep{2022ripepi}, a small sub-sample of  pure bar and disc stars could in principle be defined. All of these considerations would make it possible to study the effect of the contamination of LMC halo stars, as well as the potential impact of the warp, on the inner disc kinematics and \omp. However, these studies are beyond the scope of this Letter. The formalism to get the Cartesian  positions and velocities of stars in the LMC reference plane  from astrometric data can be found in \citet{2001vandermarel}, \citet{2002vandermarel} and \citetalias{2023jimenez}.

\begin{figure}[t]
\includegraphics[width=\columnwidth]{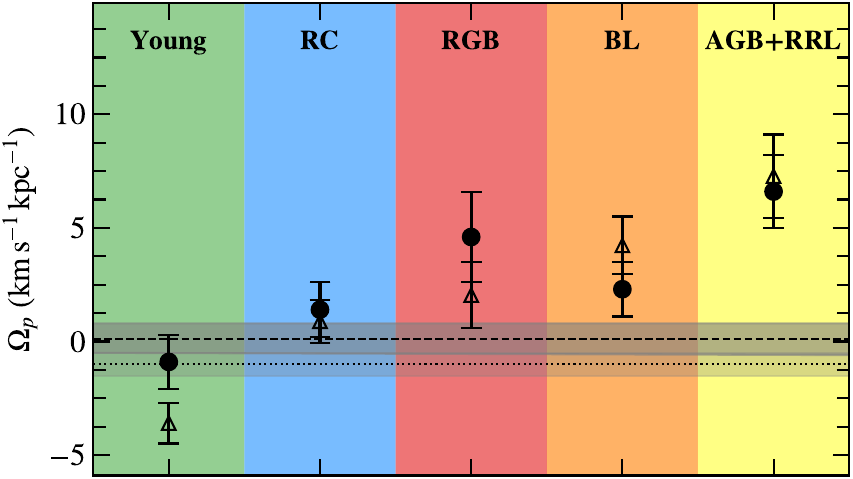}
      \caption{Modulation of $\Omega_p$ with the evolutionary phase. Filled symbols (open, respectively) are results with free (fixed) bar region. The dashed line $\Omega_p = 0.1 \pm 0.7$ \kmskpc\ is for the 
      Total sample, the dotted line $\Omega_p = -1.0 \pm 0.5$ \kmskpc\  is for  the NN Complete sample of \citetalias{2024ajimenez}.}
         \label{fig:omegap}
\end{figure}

\section{Results}
\label{sec:results}

The program \texttt{patternSpeed.py} was run with the five sub-samples of evolutionary stage, Young, RC, RGB, BL and AGB+RRL (data shown in App.~\ref{sec:app1}).  It was also run using the entire sample of $11\,535\,261$ stars combining the five phases (``Total'' sample hereafter).  We allow the program to find the bar properties without constraints on the bar region $[R_0,R_1]$. Additionally,  we measured the bar properties by fixing the  bar region to the range $R_0 \sim 0.8$ kpc, $R_1 \sim 2.3$ kpc found by \citetalias{2024ajimenez} with their NN Complete sample, enabling easier comparisons between both studies. 
The inferred pattern speeds are shown in Fig.~\ref{fig:omegap} and results are listed in Tab.~\ref{tab:res}. 
The phase angle of the LMC bar is well constrained and quite uniform throughout the stellar populations (dashed line in each map of Fig.~\ref{fig:appmapphases}). It is $\sim 19$\degr\ on average ($\sim 20$\degr\ for the fixed bar region), with a standard deviation of 3\degr. The  maximum difference of $\sim 9\degr$ occurs between the RC and Young sub-samples. 
The mean inner bound of the bar region is $R_0 \sim 0.7$ kpc, but with a total variation of 800 pc. This is caused by the outlier value for the BL sub-sample. On the contrary, the mean  bar radius is $R_1 \sim 2.3$ kpc, with very little scatter (100 pc).  

As for \omp, a first finding is that values  are very low. 
The average pattern speed is 2.2 \kmskpc\ (standard deviation of 3.6 \kmskpc), and the mean and standard deviation weighted by the number of stars are $-0.2$ and 2.8 \kmskpc, as measured within the fixed range  $[R_0,R_1]=[0.8,2.4]$ kpc for easier comparisons (right column of Tab.~\ref{tab:res}). 
The quoted uncertainties are low as well, 1.4 \kmskpc\ on average, and should be regarded as a lower limit. This corresponds to an uncertainty of more than 60\%, relatively to the mean pattern speed of 2.2 \kmskpc, however. 
Systematic uncertainties should dominate the error budget, probably at the level of a few \kmskpc. Among various sources of systematics, there are kinematical disturbances from spiral arms or the warp \citep[\citetalias{2021luri},][]{2025bjimenez}, selection effects due to sample contamination by  MW halo stars, and observational effects like the Gaia scanning law (e.g. the sawtooth pattern seen in the velocity fields, \citetalias{2021luri} and App.~\ref{sec:app2}). We think the latter 
systematics is likely to have the most significant effect on the astrometry. 

With such small angular frequencies, the LMC hosts an extremely slow bar, and this applies to each of the evolutionary phases. This is much slower than  the indirect measurement of $\Omega_p \sim 18.5$ \kmskpc\ by \citetalias{2024ajimenez} with the BV model via a corotation radius at $R_c=4.2$ kpc, but more consistent with the amplitude these authors found with the Dehnen method. 
\citetalias{2024ajimenez} suggested  that this has important implication because no corotation with disc stars   exists in this case. Indeed it is impossible to join \omp\ with the angular frequency of stars $\Omega = v_\phi/R$ given the known rotation curve(s) of the LMC.  A possibility to keep corotation in the disc  would be to have the circular velocity not well traced by the tangential velocity. For instance,  assuming the maximum value of $\sim 7$ \kmskpc\ (for the most evolved stars) and  a  corotation radius in the middle of the disc ($R_c \sim  4$ kpc), thus consistent with the BV  model of \citetalias{2024ajimenez}, it would require a circular velocity  significantly smaller than the  rotation velocity at this radius, that is, $v_{\rm {circ}} = \Omega_p R_c \sim 30$ \kms\  whereas $v_\phi \sim 80$ \kms\   (Fig. 11 of \citetalias{2024ajimenez}). Corotation radii closer to the bar radius ($R \sim 2.3$ kpc) would increase the difference with $v_\phi$ ($v_{\rm {circ}}  \sim 18$ \kms, $v_\phi \sim 60$ \kms). However, this  does not seem realistic because the opposite trend, $v_\phi < v_c$, is supposed to occur owing to  the asymmetric drift exerted on stellar populations, principally on the more evolved ones. Another possibility is that the stellar rotation curve is incorrect. This assertion would be supported  by the fact that stellar rotation curves are always measured assuming an idealized flat disc and axisymmetric motions, but the reality is quite different. Therefore, inference of a $v_\phi(R)$ profile within the framework of a  warped LMC disc like the one proposed in \citet{2025bjimenez} would certainly   help in assessing this statement. An alternative option, perhaps more realistic, is that the lack of corotation is a   transient phenomenon, as observing $\Omega_p \sim 0$ does not last long in the numerical modeling of \citet{2025jimenez}.

Even more interesting is the observation of a modulation of \omp\ with the evolutionary phase. The RC, RGB and BL phases are within $\sim 1.4-4.6$ \kmskpc, 
  the most evolved AGB+RRL phase has $\Omega_p = 6.6\pm 1.6$ \kmskpc, and the Young phase has $\Omega_p = -0.9\pm 1.2$ \kmskpc. This latter thus differs by more than 110\% from e.g. those of the RGB or the AGB+RRL phases. To our knowledge, it is the first time a modulation with the stellar phase is evidenced in a  bar.
  The different kinematics observed in the bar region through the  stellar populations is likely responsible of the different bar pattern speeds:  rotation is faster and radial motions larger for the least evolved populations (Fig.~\ref{fig:appmapphases}), whereas random motions are stronger and velocities more anisotropic for more evolved stars  (see e.g. Sect. 5.2.1 and Fig. 13 in \citetalias{2021luri}). 
  We also see that leaving the bar region free or fixed does not affect results (Fig.\ref{fig:omegap}). The phase angles and pattern speeds are barely modified, slightly increasing the difference between the least and most evolved populations. This is because it is $R_1$ that 
  has the most significant role on \omp, and this radius is uniform among the stellar phases.
Also notable is that  \omp\ and the evolution stage seem correlated, with more evolved stars  rotating faster than the bar of less evolved stars.  Further analysis will be necessary to study this trend, however.  

Then, we observe a good agreement between the result of \citetalias{2024ajimenez} and  $\Omega_p = -1.3 \pm 0.6$ \kmskpc\ for the Total sample (fixed radial range case, right column of Tab.~\ref{tab:res}). This is not surprising since both samples were built on \citetalias{2023jimenez}.  The new finding here is that $\Omega_p=-1.3$ \kmskpc\ is an equilibrium value between the phases with the largest number of stars ($-3.6$ \kmskpc\ for Young, 0.9 \kmskpc\ for RC stars) phases. This suggests significant selection effects in pattern speed derivation. The larger the size of a given sub-sample, the higher its contribution to the pattern speed measured for the Total sample. The values  obtained for our Total sample or the \citetalias{2024ajimenez}'s Complete NN sample are therefore not necessarily representative of a ``global'' LMC bar pattern speed. It is even unclear whether one should exist owing to the modulation. 

Finally, there is no reason to think that the dynamics of the LMC bar is an exception. Instead, the LMC bar kinematics behaves as expected from the viewpoint of numerical simulations, and it is very similar to that of the Milky Way bar. 
 We thus suggest that  bar pattern speed measurements, and in particular those made by means of integral field spectroscopy,  should systematically include a reference stellar population, given that such modulation with the evolution phase is likely present in many galactic bars, if not all, and that pattern speed inference is prone to selection effects. A challenge for current integral field stellar spectroscopy of galaxies, however, is its coarse spectral resolution, making the differentiation of evolutionary populations extremely difficult, if not impossible.
 
\section{Concluding remarks}
\label{sec:conclusion}
This study has measured the pattern speed of the LMC bar by means of the new Dehnen method applied to astrometric data from the \gaia\ mission for a sample of $\sim$11 million sources spanning various evolutionary phases. We find strongly different pattern speeds, and that \omp\ for the bar of less evolved stars, those along the MS, corresponds to a feature with almost no rotation whereas those of more evolved stars are faster, though still rotating at an extremely low pace. 
These results are the first insight of the modulation of \omp\ of a stellar bar with the evolutionary phase of stars. There is no singular pattern speed for the LMC bar but there are as many pattern speeds as there are stellar populations. This raises the question of whether it is relevant to reduce the bar \omp\ to a single value as stars that build it are not on the same kinematic track, and may have even experienced different evolutionary scenarios. Such low values make the LMC hosting the slowest bar ever observed in a galaxy, which confirms previous claims. No corotation radius could be found inside the disc given the LMC tangential velocity. A stationary bar, and conversely the absence of corotation is probably a transient feature provoked by the tidal interaction with the SMC, according to recent N-body simulations.
This work is the first step towards a more comprehensive analysis of the pattern speed(s) of the LMC bar as a function of the properties of its stellar populations.  The fourth release of the \gaia\ catalogue expected by the end of 2026 will provide proper motions that are at least two times more accurate than the current version. The selection of LMC disc stars will thus be improved, and the systematic effects  on \omp\ measurement significantly reduced, particularly those coming from the contamination by Milky Way halo stars and the \gaia\ scanning law. Furthermore, the 
\gaia\ Data Processing and Analysis Consortium will provide results from a detailed modeling of the spectroscopic and/or photometric data for up to a few millions of sources from our sub-samples.   
This will allow us to investigate the  modulation of the bar dynamical parameters with unprecedented details. How smooth will it be as a function of stellar metallicity and age? How does it relate to the star formation history or the recent evolution of the LMC-SMC interaction? 
Is the seemingly counter-rotating bar of the less evolved stellar phase a real feature? 
Is an \omp\ modulation present within the Milky Way bar as well? These are   questions that the \gaia\ community will unavoidably address in the near future. 

\begin{acknowledgements}
This work has made use of data from the European Space Agency
(ESA) mission \gaia\ (https://www.cosmos.esa.int/gaia), processed by
the \gaia\ Data Processing and Analysis Consortium (DPAC, https://www.
cosmos.esa.int/web/gaia/dpac/consortium). Funding for the DPAC has
been provided by national institutions, in particular the institutions participating
in the \gaia\ Multilateral Agreement. We thank Florent Renaud for insightful discussions.
       LC acknowledges financial support from the French Agence Nationale de la Recherche and  the Chilean Agencia Nacional de Investigaci\'on y Desarrollo (through the Fondo Nacional de Desarrollo Cient\'ifico y Tecnol\'ogico Regular Project 1210992. OJA acknowledges funding from ``Swedish National Space Agency 2023-00154 David Hobbs The GaiaNIR Mission'' and ``Swedish National Space Agency 2023-00137 David Hobbs The Extended Gaia Mission''. 
\end{acknowledgements}
\bibliographystyle{aa} 
\bibliography{refs.bib}
\begin{appendix}
\label{sec:app}
\onecolumn
\section{Measured bar pattern speeds}
\label{sec:app1}
In the Dehnen method, the pattern speed of the bisymmetric perturbation is given by 
$\Omega_p = (C_2\dot{S}_2 - S_2\dot{C}_2)/(2(C_2^2+S_2^2))$, 
where $C_2 = \sum_i\mu_i W(R_i) \cos 2\phi_i$ and $S_2 = \sum_i\mu_i W(R_i) \sin 2\phi_i$,  $\phi$ is the azimuthal angle in the disc and $\mu$ the individual particle mass \citep[see Appendix A of][for more details]{2023dehnen}. 
$W(R)$ is a weight  window function which allows processing the radial membership of particles to the bar. This is a major difference with the TW method whose integrals are derived along the largest radial extent possible, thus including spiral and/or outer disc regions that  do not trace at all  the bar  kinematics. A simple function delineating the bar region could thus be a top-hat filter, but \citet{2023dehnen} showed it produces a systematic overestimate of up to $25\%$ of the ground-truth \omp. \citet{2023dehnen} thus introduced a smoother transition between the bar extremity for stars beyond $R_1$ (their Eq. 25), and we used their smoothed window function as weights in our analysis.
Table~\ref{tab:res} gives the results obtained with the different stellar populations.
\begin{table*}[h]
\caption{Results of the Dehnen method applied to the LMC stellar evolutionary phases.}
\label{tab:res}      
\small       
\begin{tabular}{lrccccc|ccc}     
\hline\hline       
Sample & Number  & $R_0$ & $R_1$ & Ratio &$\phi_b$ & \omp &  Ratio &$\phi_b$ & \omp \\ 
  &  of stars & (kpc) & (kpc) & (\%) & ($\degr$) & (\kmskpc) &  (\%) & ($\degr$) & (\kmskpc) \\ 
   \hline
      Total   & $11\,535\,261$ &  1.0 & 2.3 &  28 & $19.9 \pm 0.1$ &  $0.1 \pm 0.7$ &   35 & $20.0 \pm 0.1$ &  $-1.3 \pm 0.6$\\
    \hline
   Young & $4\,500\,453$ &   1.2 & 2.3  & 24 & $23.2 \pm 0.1$ & $-0.9 \pm 1.2$ &    32   & $23.3 \pm 0.1$ & $-3.6 \pm 0.9$\\
      BL & $252\,958$ &   0.2 & 2.3 & 42 & $17.3 \pm 0.3$ & $2.3 \pm 1.2$ &  31   & $20.0 \pm 0.3$ & $4.2 \pm 1.3$\\
   RC  & $3\,737\,496$  &   1.0 & 2.2 & 29 & $14.8  \pm 0.2$ & $1.4 \pm 1.2$ &   36   &$14.8  \pm 0.2$ & $0.9 \pm 0.9$\\
   RGB & $2\,765\,725$  &   0.9 & 2.1  & 28 & $19.5 \pm 0.2$ & $4.6 \pm 2.0$ &   36   & $19.1 \pm 0.2$ & $2.0 \pm 1.4$\\
   AGB+RRL & $278\,629$ &   0.4 & 2.4 & 52 & $21.7 \pm 0.2$ & $6.6 \pm 1.6$ &   41  & $22.9 \pm 0.2$ & $7.3 \pm 1.8$\\
    \hline
   NN Complete   & $12\,116\,762$   & 0.8 & 2.3 &  35 & $20.4 \pm 0.1$ & $-1.0 \pm 0.5$\\
\hline
\end{tabular}
\tablefoot{$\phi_b$ and \omp\ are the  phase angle and pattern speed of the bar. For reference, the last row lists the results obtained by \citetalias{2024ajimenez} with their NN Complete sample.  $R_0$ and $R_1$ are the lower and outer radii of the bar region.  The left part of the Table gives results for a free bar region, the right part  for the fixed bar region $[R_0,R_1] \sim [0.8-2.3]$ kpc from \citetalias{2024ajimenez}. The ratio (in \%) is the fraction of number of stars within $[R_0,R_1]$ to the total number of stars in each sub-sample. }
\end{table*}
\section{Density and velocity maps of stellar evolutionary phases}
\label{sec:app2}
\begin{figure*}[t!]
\centering
   \includegraphics[trim=0pt 0pt 0pt 0pt, clip,height=4.5cm]{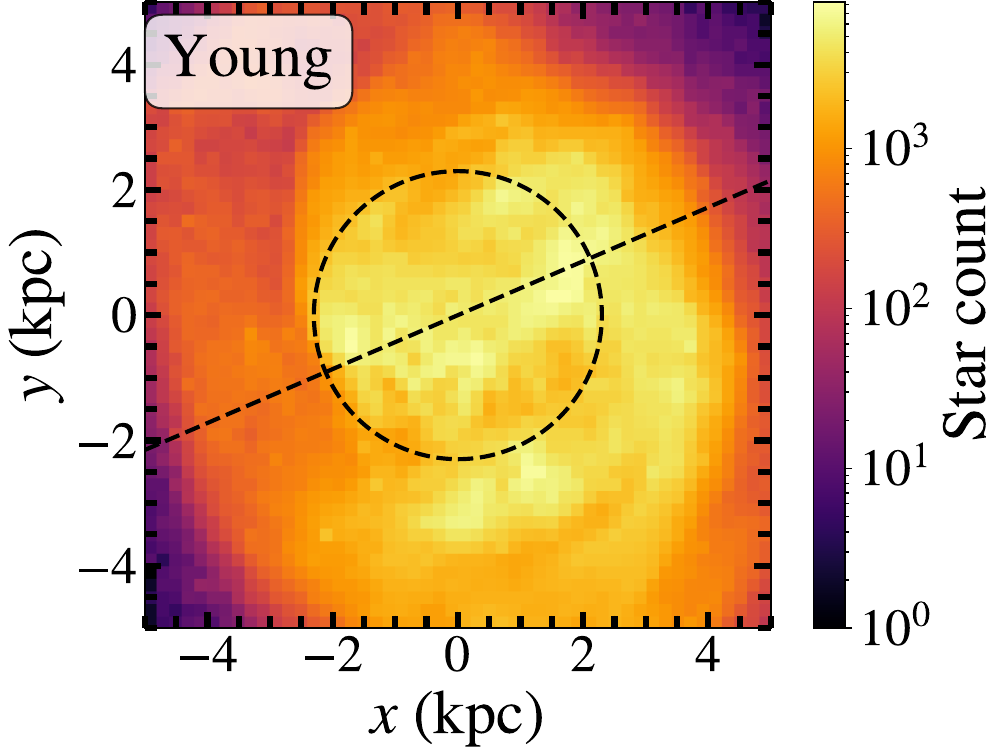}\includegraphics[trim=5pt 0pt 105pt 0pt, clip, height=4.5cm]{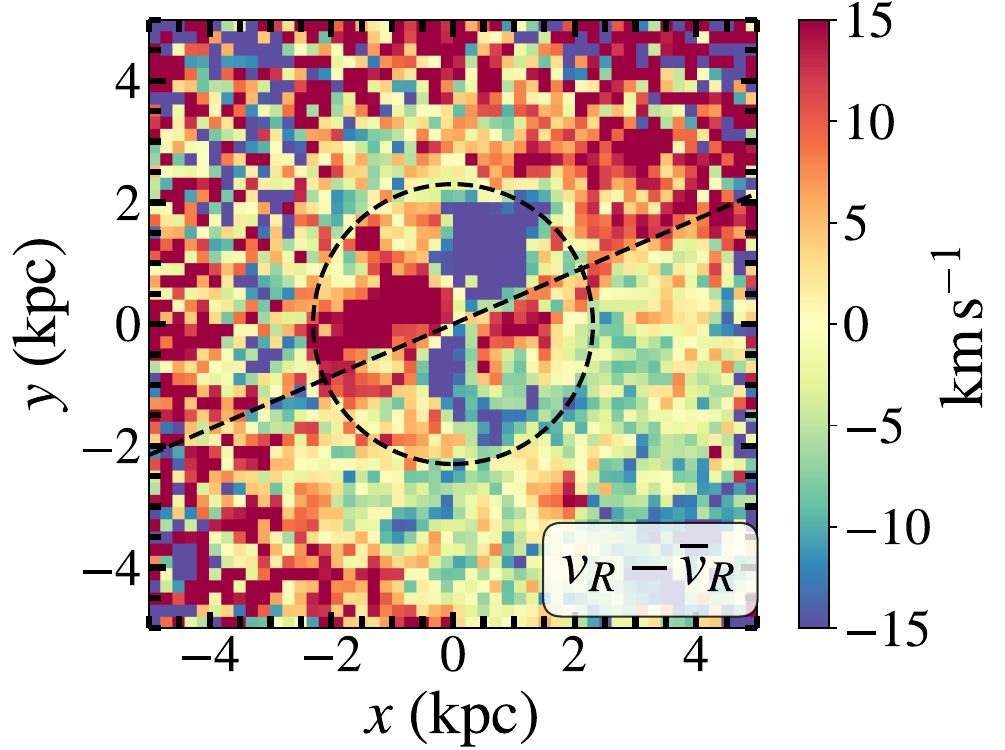}\includegraphics[trim=65pt 0pt 0pt 0pt, clip, height=4.5cm]{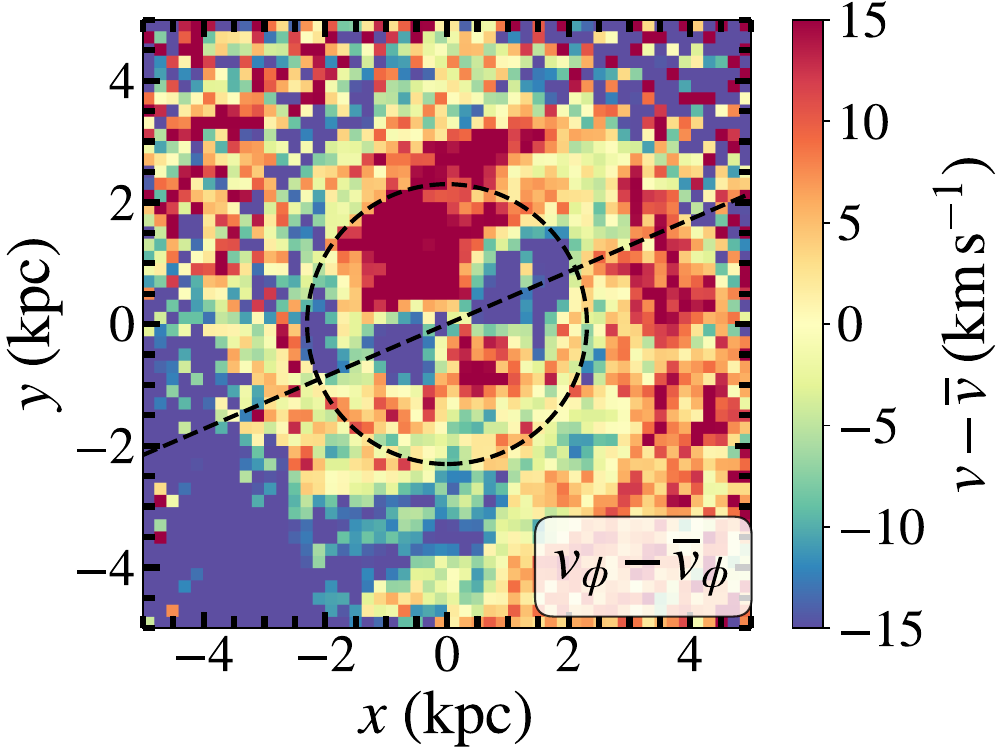}\\
   \includegraphics[trim=0pt 0pt 0pt 0pt, clip,height=4.5cm]{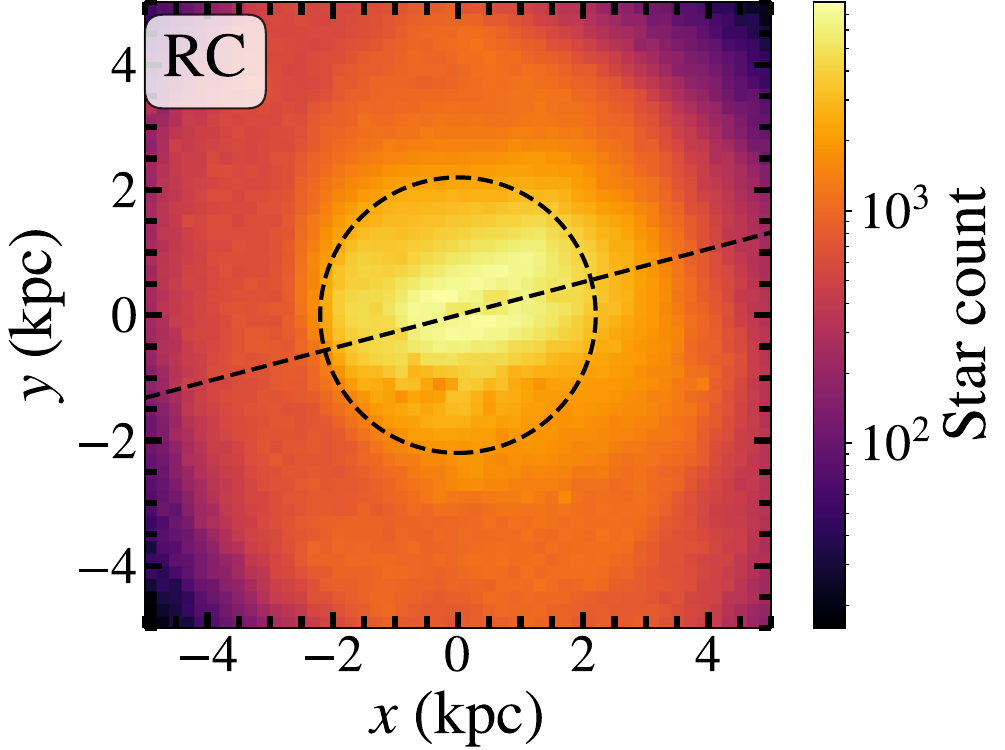}\includegraphics[trim=5pt 0pt 105pt 0pt, clip, height=4.5cm]{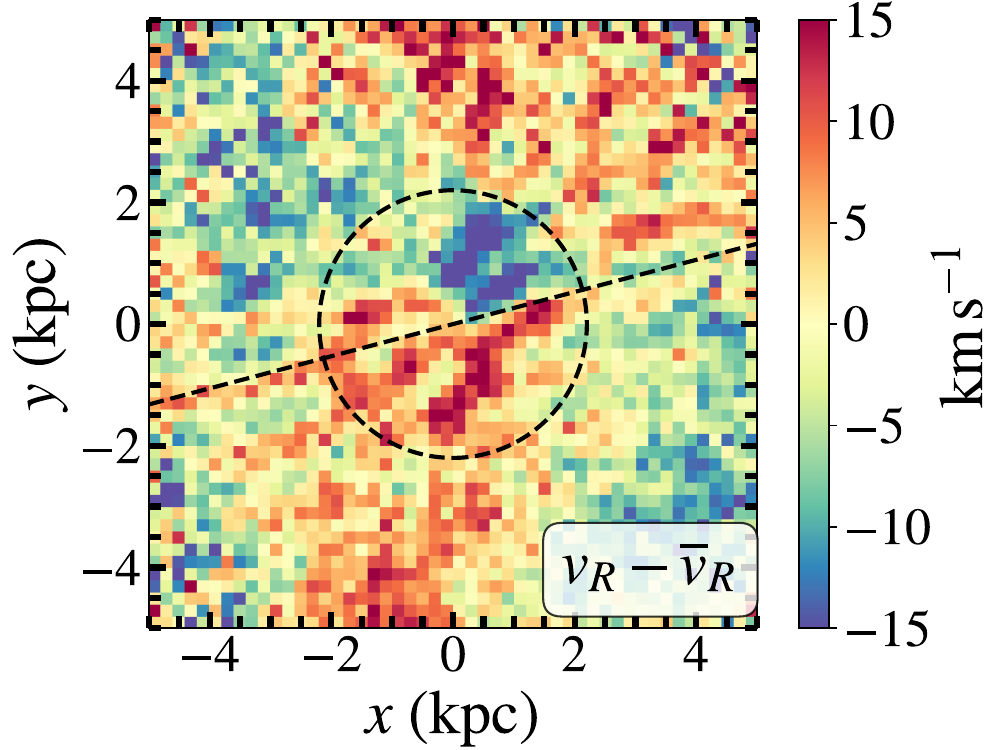}\includegraphics[trim=65pt 0pt 0pt 0pt, clip, height=4.5cm]{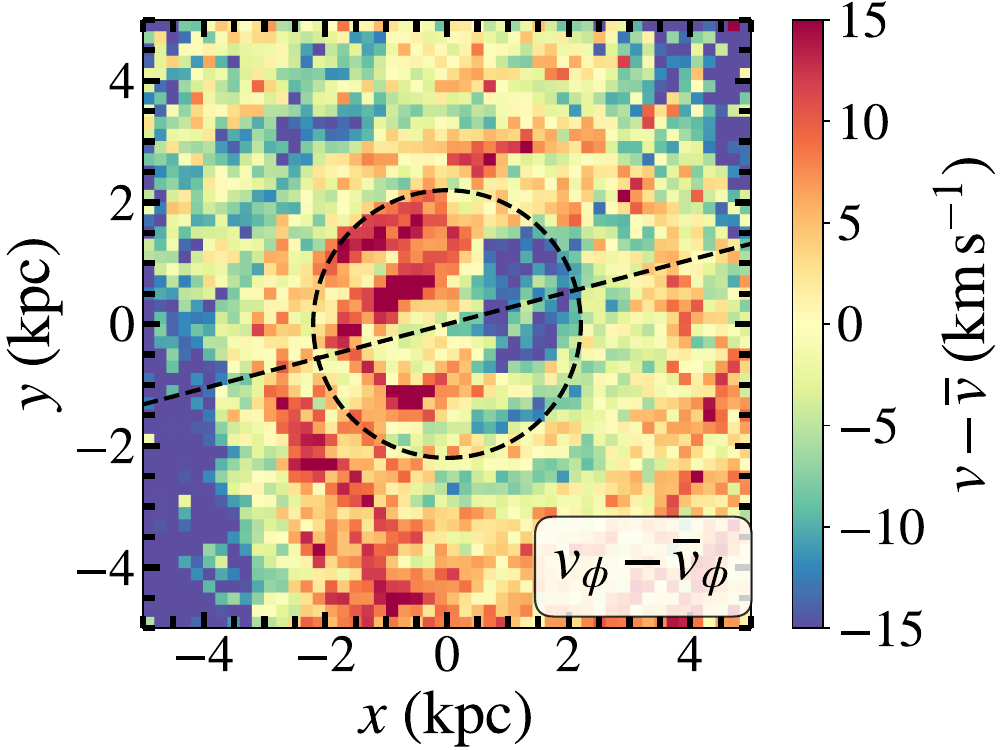}\\
   \includegraphics[trim=0pt 0pt 0pt 0pt, clip,height=4.5cm]{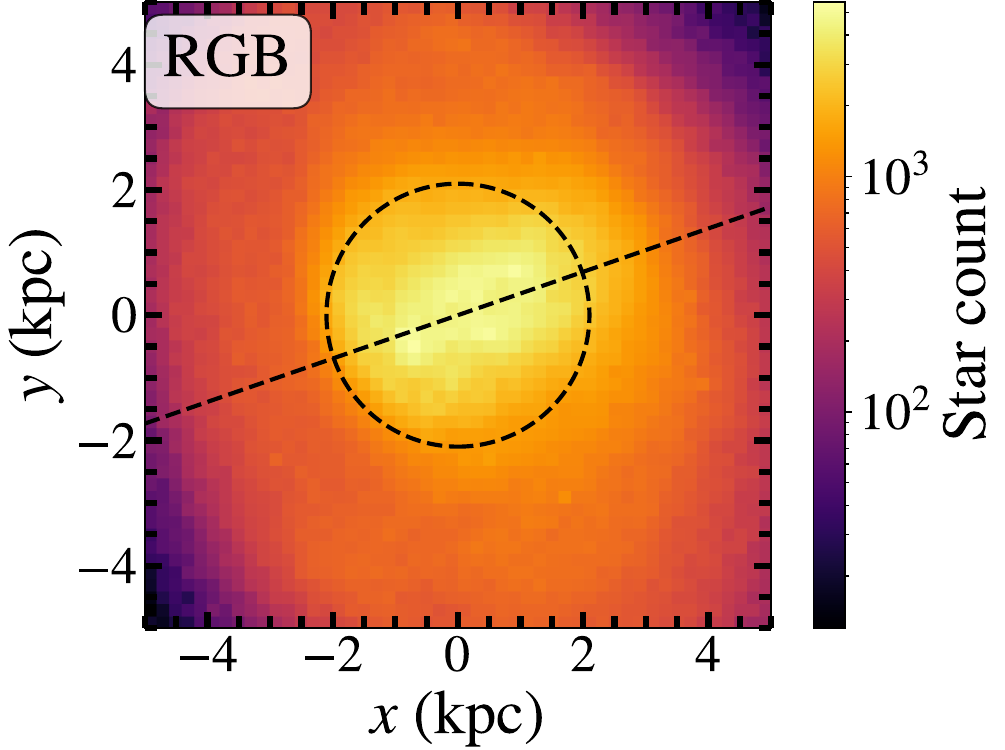}\includegraphics[trim=5pt 0pt 105pt 0pt, clip, height=4.5cm]{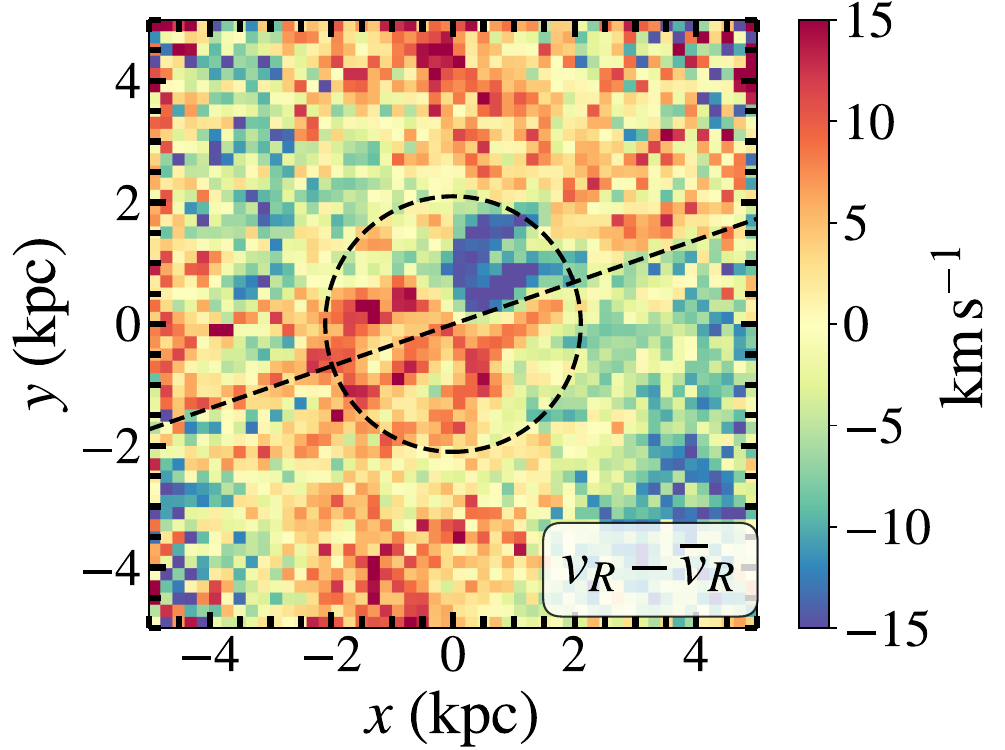}\includegraphics[trim=65pt 0pt 0pt 0pt, clip, height=4.5cm]{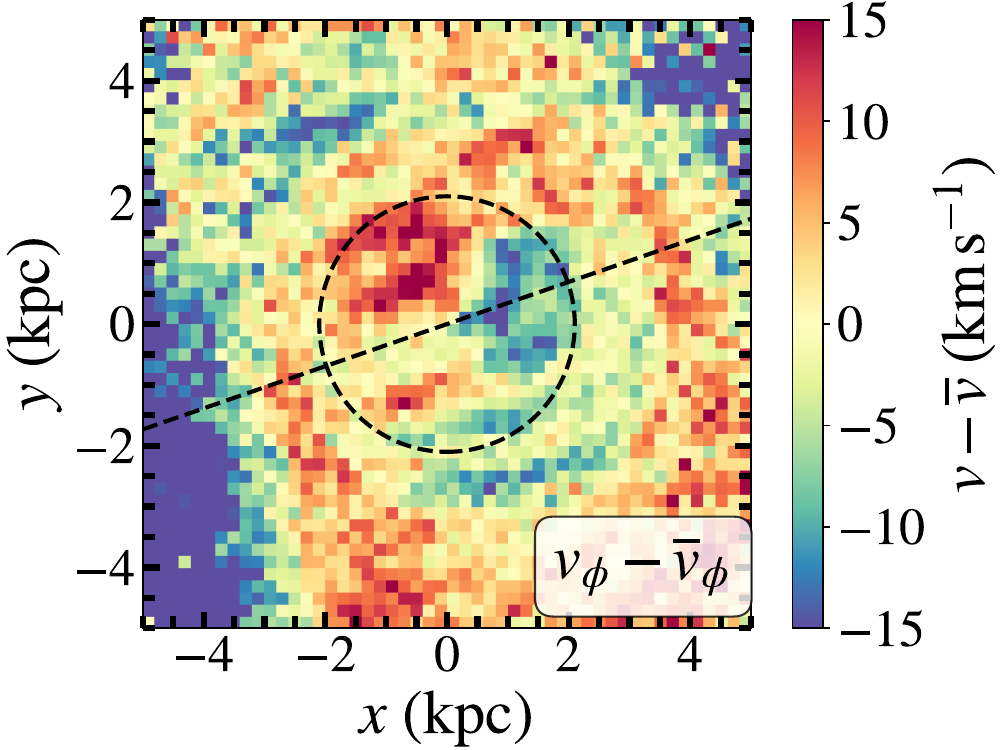}\\
   \includegraphics[trim=0pt 0pt 0pt 0pt, clip,height=4.5cm]{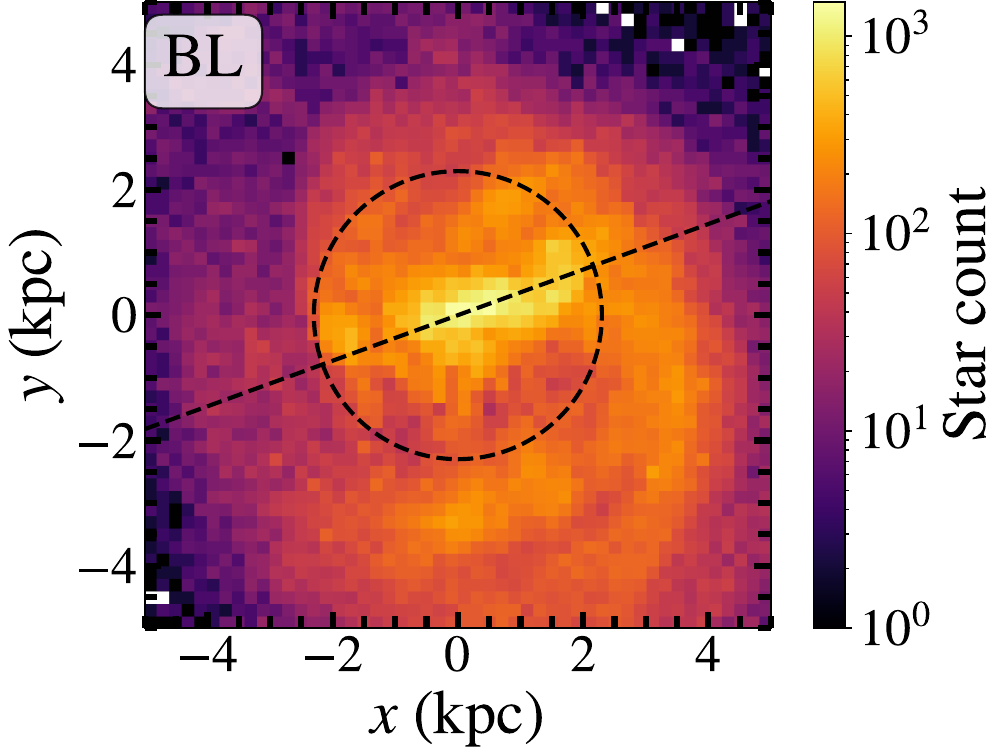}\includegraphics[trim=5pt 0pt 105pt 0pt, clip, height=4.5cm]{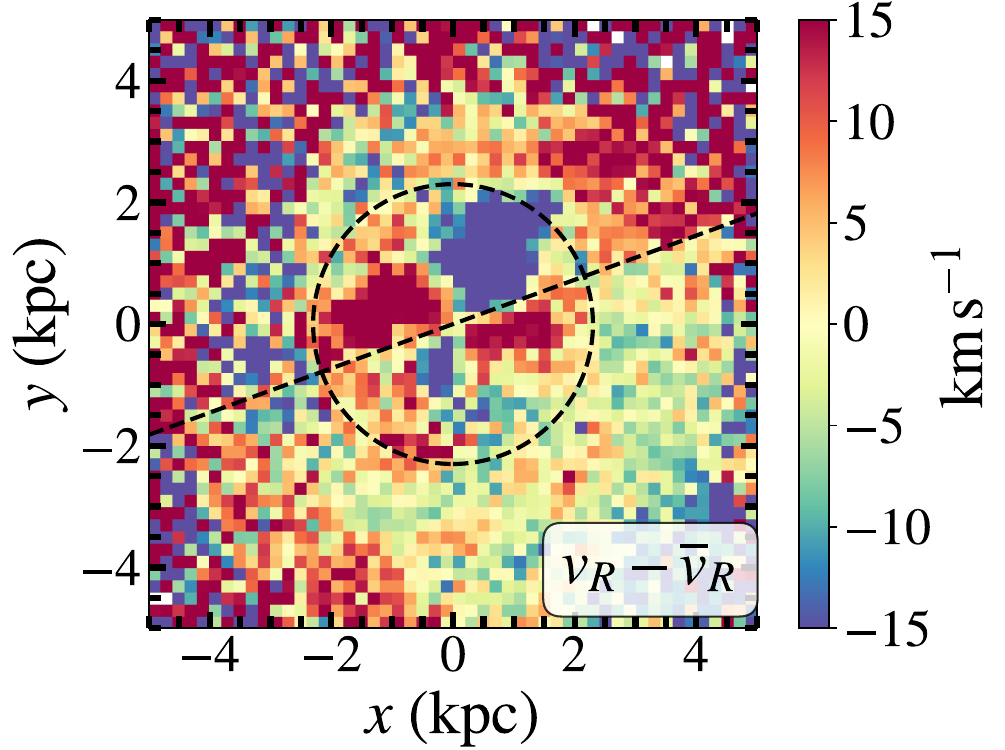}\includegraphics[trim=65pt 0pt 0pt 0pt, clip, height=4.5cm]{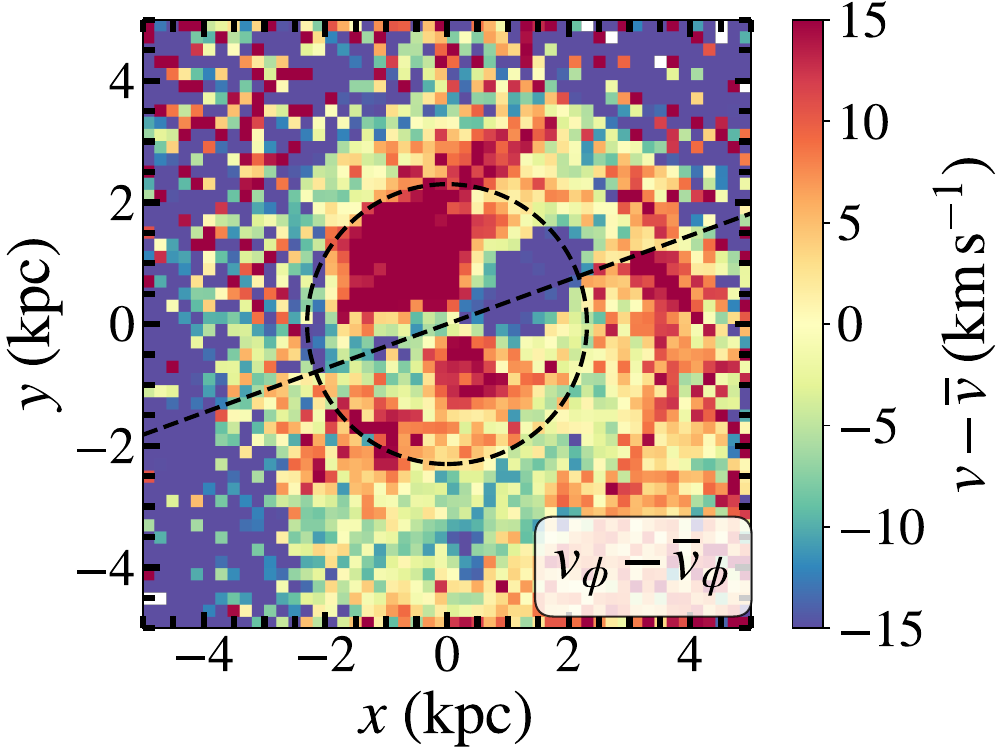}\\
  \includegraphics[trim=0pt 0pt 0pt 0pt, clip,height=4.5cm]{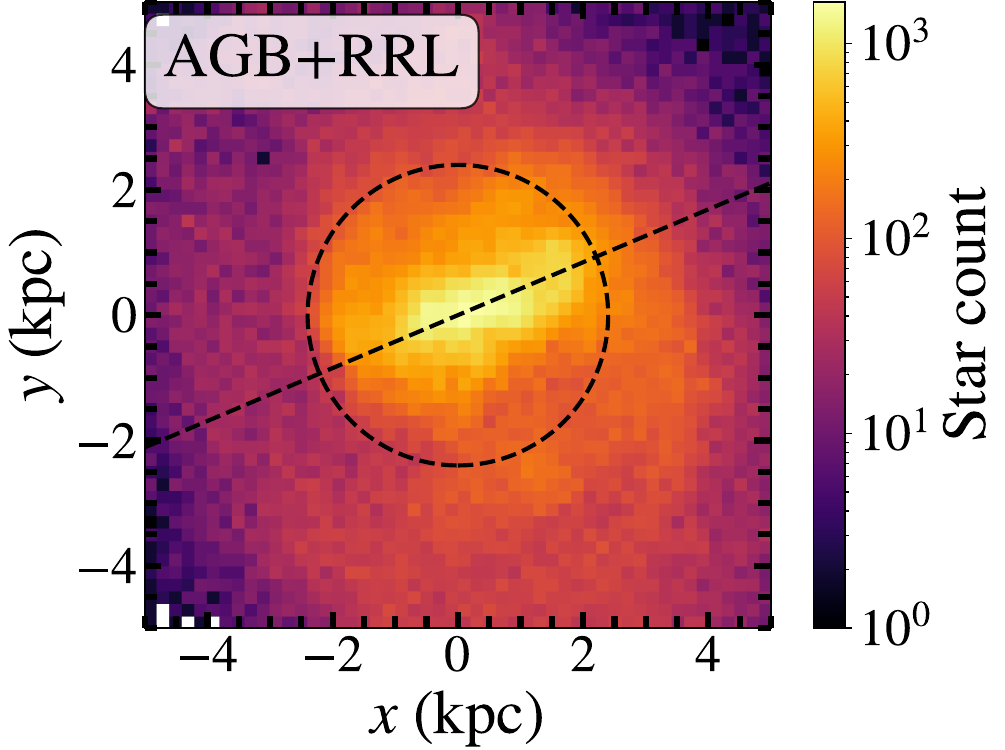}\includegraphics[trim=5pt 0pt 105pt 0pt, clip,height=4.5cm]{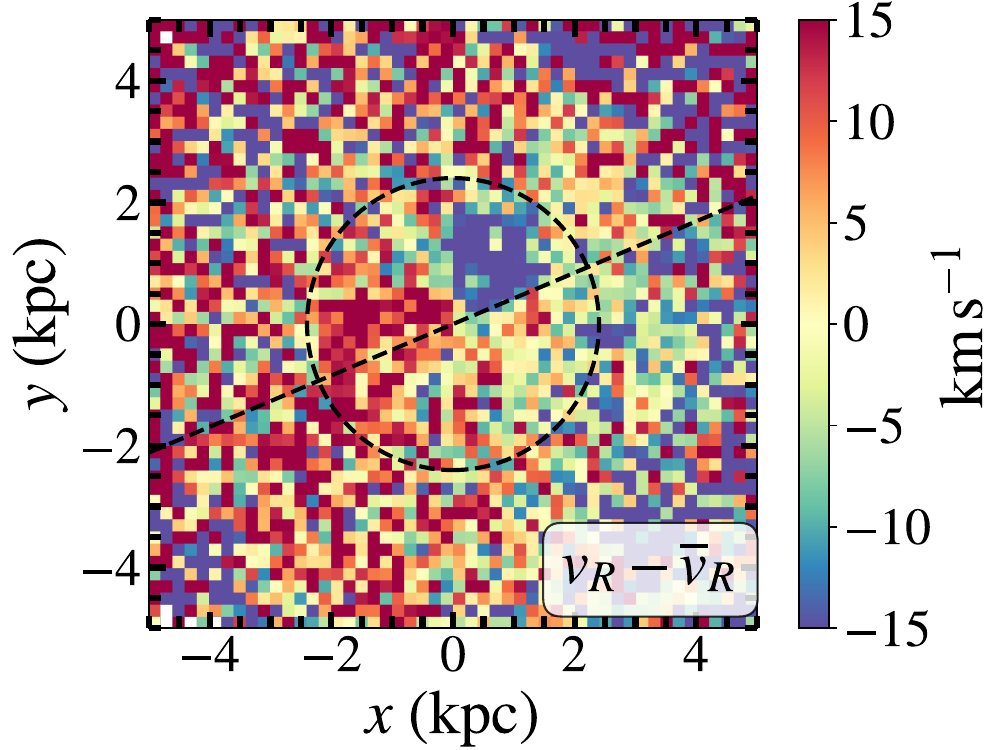}\includegraphics[trim=65pt 0pt 0pt 0pt, clip, height=4.5cm]{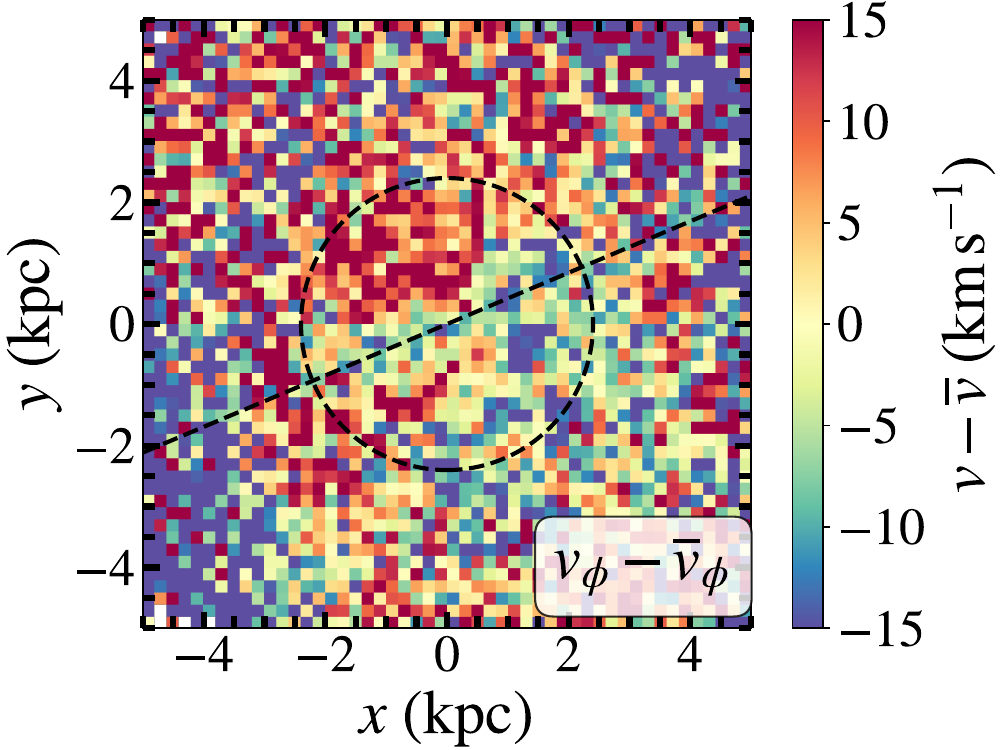}
\caption{Maps of LMC star count (left), radial ($v_R - \bar{v}_R$, middle) and tangential ($v_\phi - \bar{v}_\phi$, right) velocities for the  Young, RC, RGB, BL and AGB+RRL stellar evolutionary phases, from top to bottom respectively.  The  dashed circle  marks the bar radius (Tab.~\ref{tab:res}). A dashed line shows the direction of the bar phase angle. The $x$ and $y$ axes refer to the Cartesian coordinates in the LMC disc. The pixel scale of the grid is chosen at 200 pc.}
 \label{fig:appmapphases}
\end{figure*}

Figure~\ref{fig:appmapphases} shows the density and velocity maps for the  evolutionary phases. The radial and tangential velocities are given   relatively to their respective median value $\bar{v}$, measured within annular rings. 
 The density is higher in the inner region, where the LMC bar can be clearly identified. The Young phase shows more prominent dust lanes along the bar edges and the spiral arms than  more evolved phases. 
  All velocity maps exhibit the kinematic imprints of the stellar bar. These include sign changes of $v_R$ along directions near the major and minor axes of the bar, and lower and larger values of $v_\phi$     roughly observed inside $R_1$ along the bar major and minor axes respectively. Extrema of $v_R$ are seen near directions of $\sim 45\degr$ w.r.t. the bar major axis. They are stronger for the less evolved than the more evolved phases. The quadrupoles in the radial and tangential velocities also show some degree of asymmetry proably caused by the tidal interaction with the LMC \citep[see more details in][]{2025scholch}. Similar features have been showed in \citetalias{2021luri},  \citetalias{2023jimenez} or \citetalias{2024ajimenez}.  Also, note the stronger imprint of the sawtooth pattern caused by the \gaia\ scanning law  (\citetalias{2021luri}) in the bar region of the $v_R$ map for the RGB or RC phases, at, e.g., $(x,y) \sim (-1,-1)$ kpc.

\end{appendix}
\end{document}